\documentclass[epj,referee]{svjour2}

\usepackage{epsfig}

\def\d{{\rm d}}
\def\p{I\!\!P}

\def\beq{\begin{equation}}
\def\eeq{\end{equation}}
\def\bea{\begin{eqnarray}}
\def\eea{\end{eqnarray}}

\begin{document}

\title{Factorization Breaking in Dijet Photoproduction with a Leading
 Neutron}

\author{M.\ Klasen\inst{1,2} \and
 G.\ Kramer\inst{2}}

\institute{Laboratoire de Physique Subatomique et de Cosmologie,
 Universit\'e Joseph Fourier/CNRS-IN2P3, 53 avenue
 des Martyrs, F-38026 Grenoble, France, \email{klasen@lpsc.in2p3.fr}
 \and
 II.\ Institut f\"ur Theoretische Physik, Universit\"at Hamburg,
 Luruper Chaussee 149, D-22761 Hamburg, Germany}

\date{Date: \today}

\abstract{
The production of dijets with a leading neutron in $ep$-interactions
at HERA is calculated in leading order and next-to-leading order of
perturbative QCD using a pion-exchange model. Differential cross
sections for deep-inelastic scattering (DIS) and photoproduction are
presented as a function of several kinematic variables. By comparing
the theoretical predictions for DIS dijets to recent H1 data, the pion
flux factor together with the parton distribution functions of the
pion is determined. The dijet cross sections in photoproduction show
factorization breaking if compared to the H1 photoproduction data. The
suppression factor is $S = 0.48$ $(0.64)$ for resolved (global)
suppression.
}

\maketitle


\section{Introduction}
\label{sec:1}

In recent years, the validity of QCD factorization in hard diffractive
scattering has become an important issue experimentally and
theoretically. Factorization in hard diffraction means that the
observed cross sections in hard diffractive processes can be calculated by
a convolution of diffractive parton distributions with parton-level cross
sections. Hard-scattering factorization has been proven by Collins
\cite{1} for inclusive diffractive deep-inelastic scattering (DIS),
i.e.\ for the diffractive structure functions. It is supposed to be valid
also for subprocesses like jet production and heavy-quark production
in the DIS region. The proof of the factorization formula also appears
to be valid for the direct photoproduction of jets and heavy quarks
\cite{1}. Factorization does not hold for hard processes in diffractive
hadron-hadron scattering. Here, soft interactions between the two hadrons,
and their remnants, occur in both the initial and final state, which
prevents using the same steps as in the proof for diffractive DIS.
Therefore, factorization fails also for resolved photoproduction. The
failure of factorization in hadron-hadron scattering is observed
experimentally \cite{2}. The cross section for diffractive dijet
production at CDF is suppressed relative to the prediction based on
diffractive parton distribution functions (PDFs) from the H1
collaboration \cite{3} by one order of magnitude \cite{2}. The breaking of
hard factorization in diffractive dijet photoproduction is also firmly
established by analyses of H1 \cite{4} and ZEUS \cite{5} experimental
data. Whether these experimental data are consistent with the breaking in
the resolved component alone or whether this breaking occurs also in the
direct photoproduction cross section is still not satisfactorily proven.
It seems that the data are better described by a global suppression of the
direct and resolved contribution by about a factor of two. It is important
to note that this suppression is only visible if the data are compared to
the next-to-leading order (NLO) QCD predictions as it was first shown by
us in \cite{6,7}.
 
Factorization breaking is expected not only in the diffractive region,
$x_{\p} \ll 1$, where $x_{\p}$ is the momentum fraction transferred to the
exchanged particle in the $t$-channel, but also at larger values of
$x_{\p}$, where Regge exchanges other than the pomeron occur between
the initial and final state proton. These secondary Regge-pole
exchanges are not only present for $p \rightarrow p$ transitions but also 
in $p \rightarrow n$ transitions, in which pion exchange is strong.
Therefore dijet photoproduction with a leading neutron could also be a
candidate for factorization breaking as was already suggested in
\cite{6}. Dijet photoproduction, $e^+ +p \rightarrow e^+ + n +
{\rm jet} + {\rm jet} + X'$, with pion exchange has been studied in leading
order (LO) and NLO in \cite{8} and compared to ZEUS experimental data
\cite{9}. Recently the H1 collaboration has measured these dijet cross
sections for photoproduction ($Q^2 < 0.01$ GeV$^2$) and DIS ($Q^2 > 2$
GeV$^2$) \cite{10}. The photoproduction cross sections were compared to
the NLO predictions of \cite{8}, and good agreement was found concerning
the shape and normalization of distributions for various kinematic
variables similar to the comparison done in \cite{8} to the ZEUS data.
Neither in \cite{8} nor in \cite{10} the factorization breaking of the
dijet photoproduction cross sections has been investigated. This is
difficult, since the breaking shows up dominantly in the normalization of
the cross sections and to a much lesser extent in the shape of the 
distributions.

In the pion-exchange model, the normalization of the neutron-tagged cross
sections depends first on the splitting function of a proton into a pion
and a neutron $f_{\pi/p}(x_L,t)$. Here, $x_L$ and $t$ are the two variables 
which describe the proton-neutron vertex. $x_L$ is the fraction of the 
initial-state proton energy transferred to the neutron, and $t$ is the 
square of the momentum transfer between the proton and neutron.  Second, the
normalization depends on the parton distribution functions of
the pion, for which several models exist in the literature \cite{11}.

The pion flux $f_{\pi/p}(x_L,t)$ can, in principle, be measured in 
charge-exchange processes in soft hadronic reactions, where an initial-state 
proton is transformed into a final-state neutron, $p \rightarrow n$, with 
small momentum transfer. A successful phenomenological description of the 
corresponding data has been given in the framework of reggeized isovector 
exchanges, such as $\pi, \rho $ and $a_2$ with the pion dominating the 
$p \rightarrow n$ transition, in particular at small squared momentum
transfer $t$ between proton and neutron \cite{12}. Unfortunately these cross
sections are not given by pure Regge-exchange amplitudes. They are
modified by soft rescattering of the incoming and outgoing hadrons, which
influences the normalization of the cross sections, i.e.\ it leads to
modified flux factors $f_{\pi/p}(x_L,t)$ due to the absorption of leading 
neutrons and the ingoing protons, which depends on $x_L$ and $t$. 
Such corrections were studied some time ago by several authors \cite{13}.

The pion PDFs are constrained by di-muon and prompt-photon production data
from fixed target experiments that are sensitive to the valence quark
distribution in a Bjorken-$x$ range relevant for dijet production on pions
\cite{11}. Unfortunately, these constraints are not very restrictive so that
the existing parameterizations \cite{11} differ in the relevant $x$-range,
which leads to appreciable differences in the calculated dijet
photoproduction cross sections \cite{8}.

The most direct way to determine the flux factor times the structure
function of the pion $F_2^{\pi}(x,Q^2)$ is to measure the inclusive DIS cross
section with a tagged neutron. On the basis of Collins' factorization 
proof, we expect in this case no absorptive interactions. Such cross sections 
for leading neutrons have been measured in $ep$-collisions at HERA \cite{14}.
Unfortunately these data have not been analyzed towards determining the
PDFs of the pion, assuming a fixed ansatz for the $p \rightarrow n$ vertex, as 
it has been done towards the  PDFs of the pomeron from diffractive 
semi-inclusive cross sections \cite{3,15}. Therefore we shall follow a 
different route. Assuming one of the PDFs of the pion in the literature, for
example those of Gl\"uck, Reya and Vogt (GRV) \cite{11}, we determine the
normalization of $f_{\pi/p}(x_L,t)$  by comparing the NLO dijet cross sections 
for $Q^2 > 2$ GeV$^2$ to the data of the H1 collaboration \cite{10}. With this 
$f_{\pi/p}(x_L,t)$  in conjunction with GRV's pion PDFs, we have calculated 
the NLO dijet cross section for photoproduction ($Q^2 \simeq 0$)
on the basis of \cite{8} and compare them with the experimental results in
\cite{10}. Here the experimental tagged neutron DIS and photoproduction cross 
sections are obtained in a common analysis in the same region of the 
neutron kinematic variables.
If we assume that the dijet production data in the DIS region are not
influenced (or influenced very little) by absorptive corrections, i.e.\
have no (or only a very small) breaking of factorization, then the failure
of our NLO photoproduction cross sections to describe the corresponding
data of \cite{10} will give us the amount of factorization breaking for
photoproduction.

In Section 2, we shall describe the calculation of the dijet cross sections
with a leading neutron, together with the kinematic variables, and define
our input for the pion flux and the pion PDFs. The NLO cross sections are
compared to several measured DIS cross sections from \cite{10}, so that
our assumptions concerning $f_{\pi/p}(x_L,t)$ and the pion PDFs can be tested. 
Section 3 contains our results for the dijet cross sections in the 
$Q^2 \simeq 0$ region and the comparison with the experimental data of 
\cite{10}. On this basis we test also whether the factorization breaking can 
be attributed to the resolved component alone. In Section 4, we give a short 
summary and draw some conclusions.

\section{Dijet Cross Sections in DIS}
\label{sec:2}

\subsection{Kinematic Variables and Input}

The event kinematics has already been described in \cite{10}. Here we
recall the definition of those variables which are needed in the
calculations of the cross sections. The reaction
\begin{equation}
 e^{+}(k) + p(P) \rightarrow e^{+}(k') + n(P') + X(p_X),
\end{equation}
where $X$ is the hadronic system containing at least two jets, is
characterized by the four-momenta $k$ and $k'$ of the initial and scattered
positron and by $P$ and $P'$, the four-momenta of the ingoing proton and
outgoing neutron, respectively, as sketched in Fig.\ 1 (left). The positron-photon
%
\begin{figure}
 \centering
 \includegraphics[width=0.49\columnwidth]{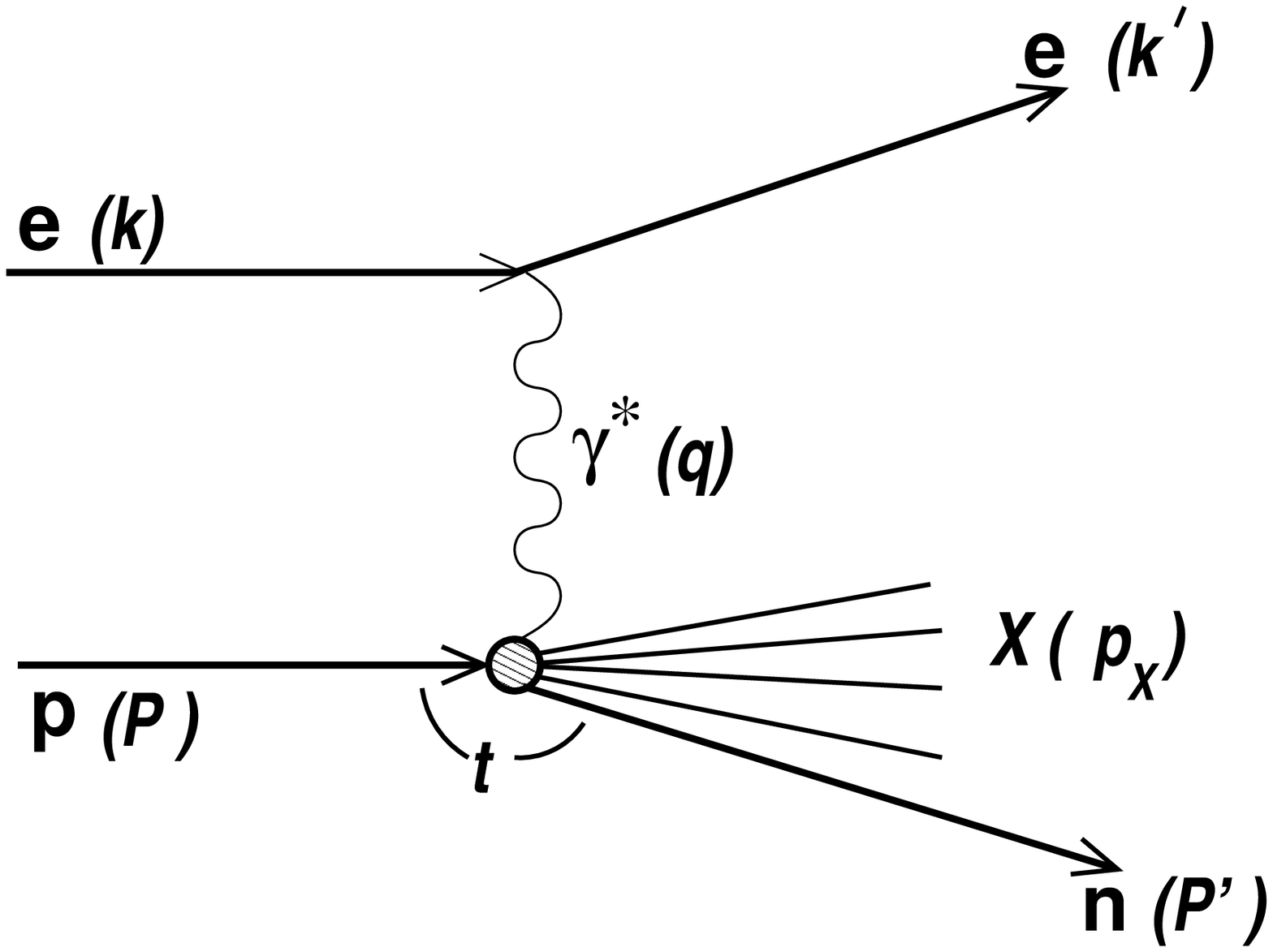}
 \includegraphics[width=0.49\columnwidth]{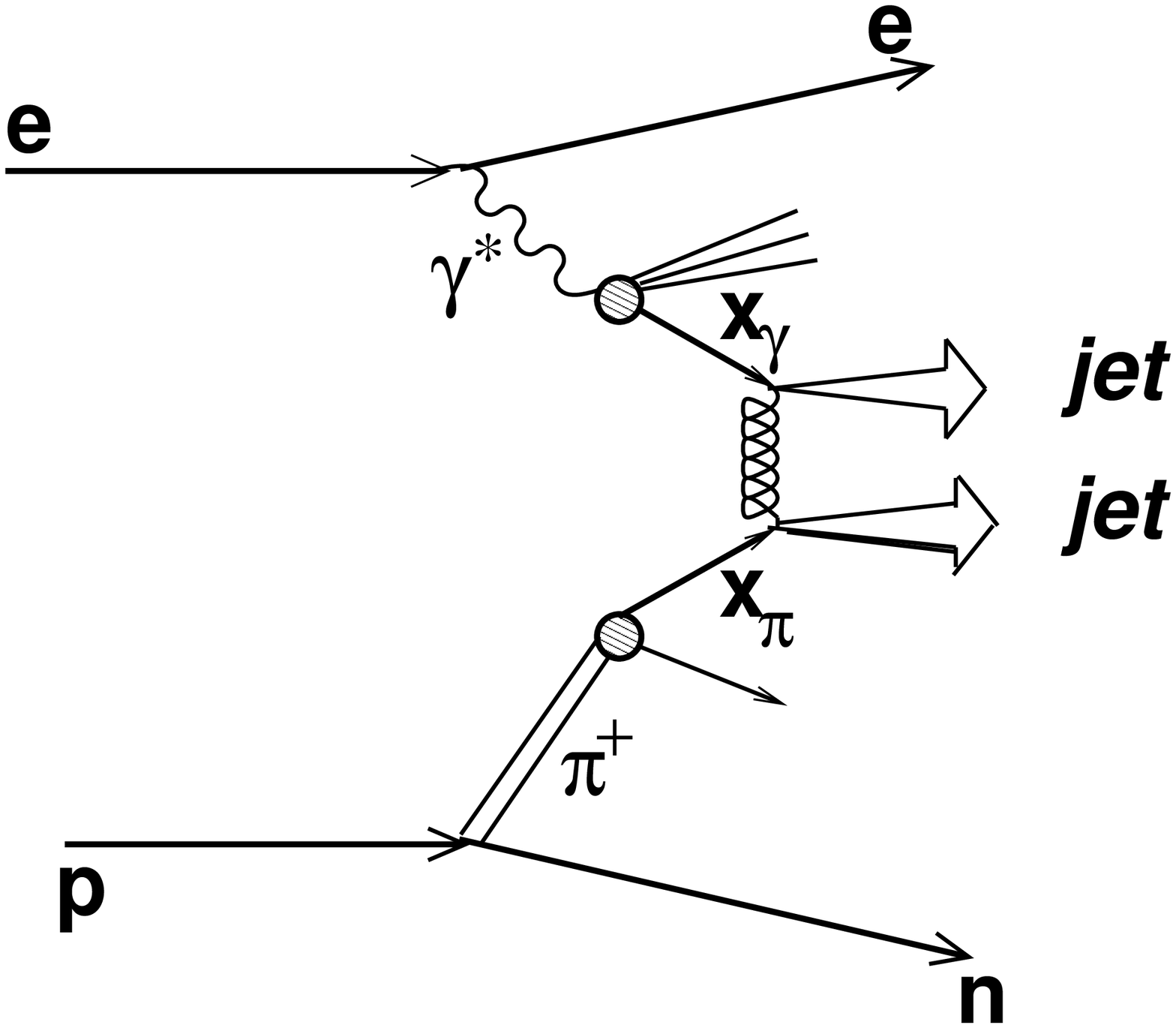}
 \caption{\label{fig:1}Generic Feynman diagrams for the scattering process
 $e+p\to e+n+X$ (left) and for the production of two jets in the one-pion
 exchange model (right).}
\end{figure}
%
vertex is described by the exchanged photon virtuality $Q^2$ and the
positron's inelasticity $y$,
\bea
   Q^2 = -q^2 = -(k-k')^2 &~~~{\rm and}~~~&  y=\frac{Pq}{Pk}.
\eea
In the H1 experiment \cite{10}, $2<Q^2<80$ GeV$^2$ and $0.1<y<0.7$ in 
the DIS region and $Q^2 < 10^{-2}$ GeV$^2$ and $0.3<y<0.65$ for the
photoproduction selection. The protons at HERA have the energy $E_p=820$ GeV
and collide with $E_e=27.6$ GeV positrons.

The two variables, which describe the proton-neutron vertex, are the fraction
$x_L$ of the energy of the initial-state proton $E_p$ carried by
the neutron and the square of the momentum transfer $t$ between the proton
and the produced neutron,
\begin{eqnarray}
 x_L &=& \frac{P'q}{Pq} \simeq \frac{E_n}{E_p} ~~{\rm and} \nonumber\\
 t &=& (P-P')^2 \simeq \frac{-p_{T_n}^2}{x_L} -(1-x_L)
 \left(\frac{m_n^2}{x_L}-m_p^2\right),
\end{eqnarray}
where $E_n$ is the neutron energy, $p_{T_n}$ is the momentum component of the 
neutron transverse to the direction of the ingoing proton, and $m_n$ and $m_p$
are the neutron and proton masses, respectively. $x_L$ and $t$ are determined
from the measured energy and scattering angle of the leading neutron. In the H1
experiment $E_n > 500$ GeV and $\theta _n < 0.8$ mrad.

In the pion-exchange model, the photon interacts with a pion emitted from the
proton. In this model, the process $e^+ + p \rightarrow e^+ +n+ {\rm jet} +
{\rm jet} +X'$
as sketched in Fig.\ 1 (right) is described by the variable $x_{\pi}$, which, 
neglecting masses, is the fraction of the four-momentum of the pion 
$q' =P-P'$ participating in the hard interaction. It is related to $x_p$, the 
fraction of the four-momentum of the proton, which enters the hard interaction,
by $x_p=x_{\pi}(1-x_L)$. $x_L$ is related to $x_{\p}$ introduced in Section 1 in
connection with diffractive jet production via $x_L=1-x_{\p}$.

In the scattering of $2 \rightarrow 2$ massless partons, the fractions
of the four-momenta $q = k-k'$ and $q' = P-P'$ transferred to the partons are
given by
\beq
   x_{\gamma}^{\rm jet} = \frac{\sum_{\rm jets} E_T^{\rm jet} e^{-\eta^{\rm jet}}}{2yE_e}
\eeq
and
\beq
   x_{\pi}^{\rm jet} = \frac{\sum_{\rm jets} E_T^{\rm jet} e^{\eta^{\rm jet}}}{2E_p(1-x_L)},
\eeq
so that
\begin{equation}  
  x_p^{\rm jet} = \frac{\sum_{\rm jets} E_T^{\rm jet} e^{\eta^{\rm jet}}}{2E_p}.
\end{equation}
The sums in Eqs.\ (4)-(6) run over the variables of the two jets in the final
state. $E_T^{\rm jet}$ and $\eta^{\rm jet}$ denote the transverse energies and pseudorapidities
in the laboratory system. Strictly speaking, Eq.\ (4) is correct only for
photoproduction, where $q^2\simeq0$, and furthermore $q'^2=0$. The energy
fraction contributing by 
the exchanged virtual photon to the production of the dijets is $x_{\gamma}^{\rm jet}$, 
whereas the corresponding contribution of the virtual pion (or possibly of a 
reggeized $\rho$ or $a_2$) is $x_{\pi}^{\rm jet}$. In Eq.\ (4), $E_{\gamma}=yE_e$ is the 
energy of the ingoing virtual photon. In NLO, also three jets can be produced 
in the final state, but these contributions have been removed from the
theoretical prediction in accordance with the experimental analysis, which
contains only an exclusive dijet sample (see below).

As in the jet analysis of the experimental data \cite{10}, we use the cone
algorithm with radius $R'=1$ \cite{16} for the jet definition and the
combination of two partons into one jet in the NLO contributions of the DIS
and the photoproduction sample. In our previous work \cite{8}, we had
used the $k_T$-cluster algorithm instead. The jet finding is performed in the 
$\gamma ^{*} p$ center-of-mass frame with transverse energies defined 
relative to the $\gamma ^{*}$ momentum axis. The laboratory pseudorapidity of 
each jet is restricted to the region $-1 < \eta^{\rm jet} < 2$. The transverse 
energies of the two jets with the largest $E_T$ are constrained to the region
$E_T^{\rm jet1} > 7$ GeV and $E_T^{\rm jet2} > 6$ GeV for both the DIS and
the photoproduced jets in accordance with the requirement of asymmetric
$E_T^{\rm jet}$-cuts \cite{16a}. The calculated cross sections are restricted
to the selection of exactly two jets, i.e.\ the rather small contribution of
three jets with $E_T^{\rm jet3} > 6$ GeV is not included as in the experimental
selection  \cite{10}.

For the production of dijets in the DIS region, we consider only the
contribution of the directly coupled $\gamma^{*}$, although at $Q^2$ as 
low as $2$ GeV$^2$ the resolved contribution might be relevant
\cite{16b}. For the 
photoproduction of dijets we take into account the direct and the resolved 
process both in LO and NLO. The observable $x_{\gamma}^{\rm jet}$ is sensitive to the 
amount of direct and resolved processes.

In the pion-exchange model, the cross section for $\gamma ^{*}p$ scattering to
the final state $nX$ (see Eq.\ (1)) takes the form
\begin{equation}
 \d\sigma(\gamma^{*}p \rightarrow nX) = f_{\pi/p}(x_L,t)~
 \d\sigma(\gamma^{*}\pi^+ \rightarrow X),
\end{equation}
where $f_{\pi/p}$ is the pion flux for the transition $p \rightarrow n+\pi^+$
and  $\d\sigma(\gamma^{*} \pi^+ \rightarrow X)$ stands for the cross
section of the hard $\gamma^{*}$-$\pi^+$ interaction. The splitting function or
pion flux is usually parameterized by different forms. We choose
\begin{equation}
 f_{\pi/p}(x_L,t) = \frac{1}{4\pi}\frac{g^2_{n\pi p}}{4\pi}
 \frac{-t}{(m_{\pi}^2-t)^2} (1-x_L)^{1-2\alpha_{\pi}(t)} [F(x_L,t)]^2.
\end{equation}
Here, $g_{n\pi p}$ is the coupling constant of the $n\pi p$ vertex, $m_{\pi}$ 
is  the pion mass and  $\alpha _{\pi}(t) = \alpha '(t-m_{\pi}^2)$ is the Regge 
trajectory of the pion. $F(x_L,t)$ is a form factor, which describes the
off-shell behavior of the virtual pion and/or possible final-state
rescattering effects of the neutron. We choose the so-called light-cone
form factor
\begin{equation}
 F(x_L,t) = \exp[R^2(t-m_{\pi}^2)/(1-x_L)].
\end{equation}
This choice is usually associated with the flux without Regge trajectory
factor, i.e.\ $\alpha'=0$ in Eq.\ (8) \cite{17}. The pion-nucleon coupling
constant is known from low-energy $\pi N$ and $NN$ scattering data. We take 
$g^2_{n\pi p}/4\pi=2\times 14.11$ \cite{18}. Other choices of $F(x_L,t)$ have 
been used in the literature. We consider only the form in Eq.\ (9), since it
has been  used also in \cite{10}.

Another important input are the PDFs of the pion. For this several
choices are available in the literature \cite{11}. We choose the
parameterization of Gl\"uck, Reya and Vogt (GRV) \cite{11} which was also 
chosen in \cite{10}. They provide PDFs in LO and NLO and also include the 
charm contribution. For the calculation of the photoproduction cross sections 
we need the PDFs of the photon for the resolved part. A popular 
parameterization is the one of Gl\"uck, Reya and Vogt (GRV) \cite{19} which
was also the choice in \cite{10}. The $\Lambda_{\overline{MS}}$ parameter,
which we need in the LO and NLO formulas for $\alpha_s$, is adjusted to the
PDG 2006 edition \cite{20} value  for $\alpha_s(m_Z)=0.1176$. This yields for 
four flavors $\Lambda^{(4)}_{\overline{MS}} =0.118$ GeV in LO and $\Lambda
^{(4)}_{\overline{MS}}=0.307$ GeV in NLO. Of course, our NLO predictions
for DIS and photoproduction dijet cross sections depend on this value. For the 
LO (NLO) predictions we use LO (NLO) hard scattering matrix elements with the 
one-loop (two-loop) formula for $\alpha_s$ and the 
$\Lambda^{(4)}_{\overline{MS}}$ values given above. Unfortunately, the 
$\Lambda$-values used for the evolution of the GRV pion and photon PDFs are
somewhat different, namely $0.200$ GeV in LO and NLO. In the case of
photoproduction, the photon flux is 
calculated with the usual Weizs\"acker-Williams approximation including the 
non-logarithmic corrections as calculated in \cite{21}. The renormalization 
and factorization scales are equal to the maximum transverse energy of the 
outgoing jets. Now, all parameters and PDFs are specified except for the 
radius $R$ in the light-cone form factor in Eq.\ (9). The value of this parameter will 
be fixed by comparison of the theoretical cross section in DIS dijet production
with the data of \cite{10}, i.e.\ with the measured cross sections 
$\d\sigma/\d E_T^{\rm jet}$, $\d\sigma/\d\eta^{\rm jet}$, 
$\d\sigma/\d\log_{10}(x_{\pi}^{\rm jet})$, and $\d\sigma/\d Q^2$.
In \cite{10}, the  photoproduction dijet data have been described very well
with the choice  $R=0.65$ GeV$^{-1}$.

In addition to pion exchange, secondary Regge exchanges (as for
example $\rho$ and $a_2$) are possible. Such contributions could be
disentangled by a careful study of the flux factor as a function of
$x_L$ and $t$ (or $p_{T_n}$). Such data are not available. In \cite{10},
the neutron energy dependence of the DIS and photoproduction dijet
events were measured. The uncorrected data were compared to a
pion-exchange model within a Monte Carlo simulation that included
detector effects. The photoproduction data are reasonably described in
shape and magnitude, whereas the DIS data are reproduced quite well in
shape, but somewhat overestimated in magnitude. We take these results
for a good indication that pion-exchange is dominating over $\rho$-
and $a_2$-exchanges at least in the region of $x_L>0.6$. Since
additional $\rho$- and $a_2$-exchanges influence only the $p\to n$
flux factor, this factor should be the same for the DIS and
photoproduced dijets due to factorization.

\subsection{DIS Dijet Cross Sections}

The calculations of the dijet cross sections in the DIS region have been 
performed
with the NLO Monte Carlo program JETVIP \cite{22}. This program calculates jet 
cross sections in LO and NLO in DIS using the so-called phase space slicing 
method with an invariant mass cut-off to cancel the infrared and collinear 
singularities. The program, originally constructed for jet production in 
$\gamma ^{*}p$ interactions, could easily be modified for our purpose to 
calculate jet cross sections with a tagged neutron. The only change is that 
the PDFs of the proton are replaced by the PDFs of the pion times the pion 
flux.

Since the neutron kinematics is not fixed in detail by the experiment, we had
to integrate over a finite region in $x_L$ and $t$ according to Eqs.\ (3).
We did this in accordance
with the specifications of the H1 experimental analysis. Except for the 
outgoing positron and the leading neutron, the final state consists of two or
three jets in addition to the remnant jet of the pion. The two-jet sample
contains the bare parton jets from the LO and virtual NLO contributions
and the two jets originating from the recombination of two partons in the 
three-parton contribution using the cone algorithm. Then we calculated the
differential cross section $\d^2\sigma/\d E_T\d\eta$. where $E_T$ and $\eta$
are the transverse energy and the rapidity of the jets in the two- or three-jet
sample with $E_T^{\rm jet1} > 7$ GeV and $E_T^{\rm jet2} > 6$ GeV and 
$E_T^{\rm jet3} <6$ GeV, i.e.\ from the three-jet sample the events with a
hard jet with $E_T^{\rm jet3} > 6$ GeV are left out. This defines the
exclusive dijet sample
as in the analysis by H1 \cite{10}. The cross section $\d\sigma/\d E_T\d\eta$
was integrated over the region $-1 < \eta^{\rm jet1,2} < 2$ to yield
$\d\sigma/\d E_T$ and integrated over $E_T$ with the asymmetric $E_T$-cut
defined above to give us $\d\sigma/\d\eta^{\rm jet}$ and similarly for
$\d\sigma/\d\log_{10}(x_{\pi}^{\rm jet})$ and $\d\sigma/\d Q^2$.

The results for $\d\sigma/\d E_T^{\rm jet}$, $\d\sigma/\d\eta^{\rm jet}$,
$\d\sigma/\d\log_{10}(x_{\pi}^{\rm jet})$ and $\d\sigma/\d Q^2$ are
shown in Figs.\ 2a-d. 
%
\begin{figure*}
 \centering
 \includegraphics[width=0.49\columnwidth]{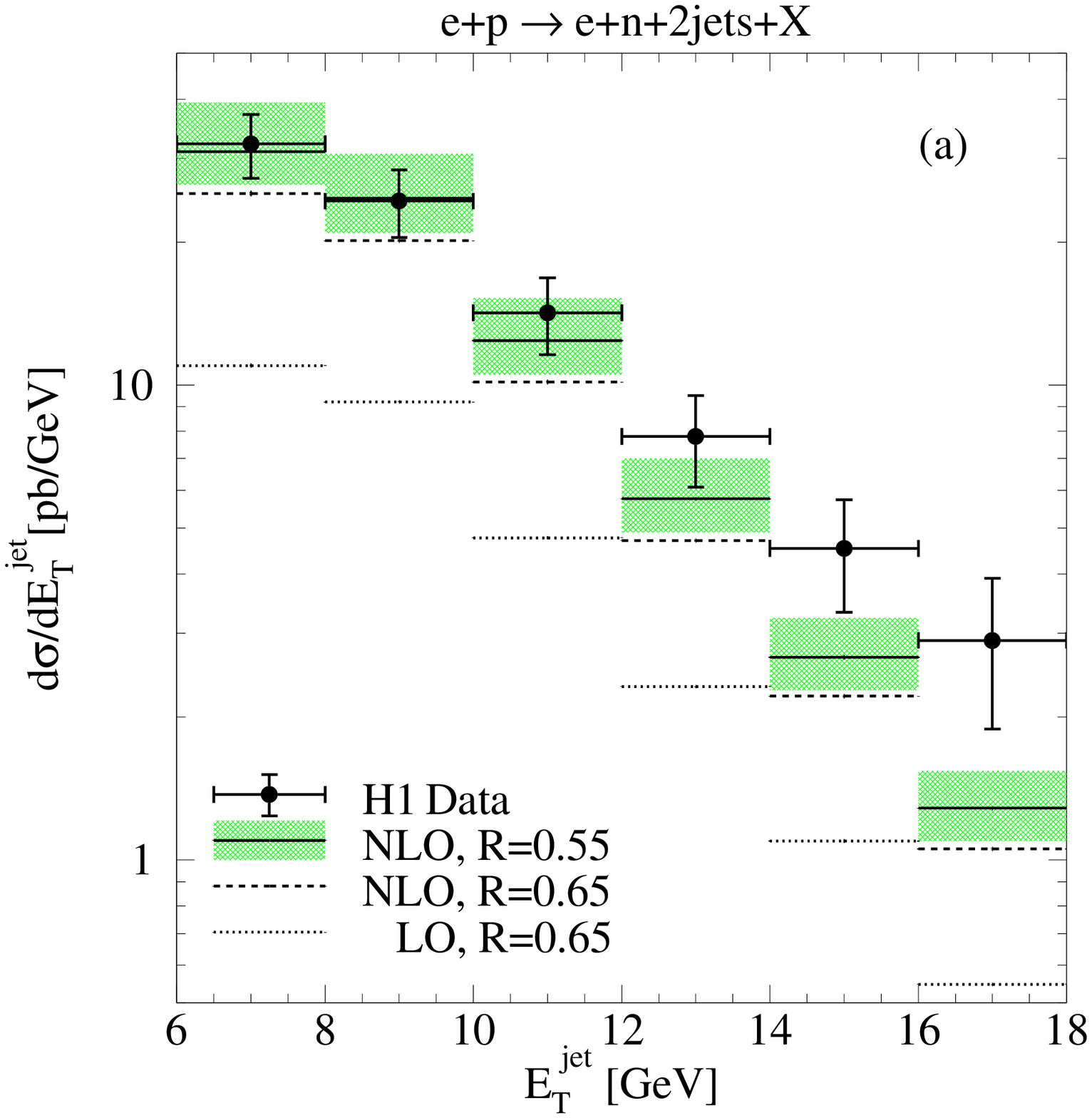}
 \includegraphics[width=0.49\columnwidth]{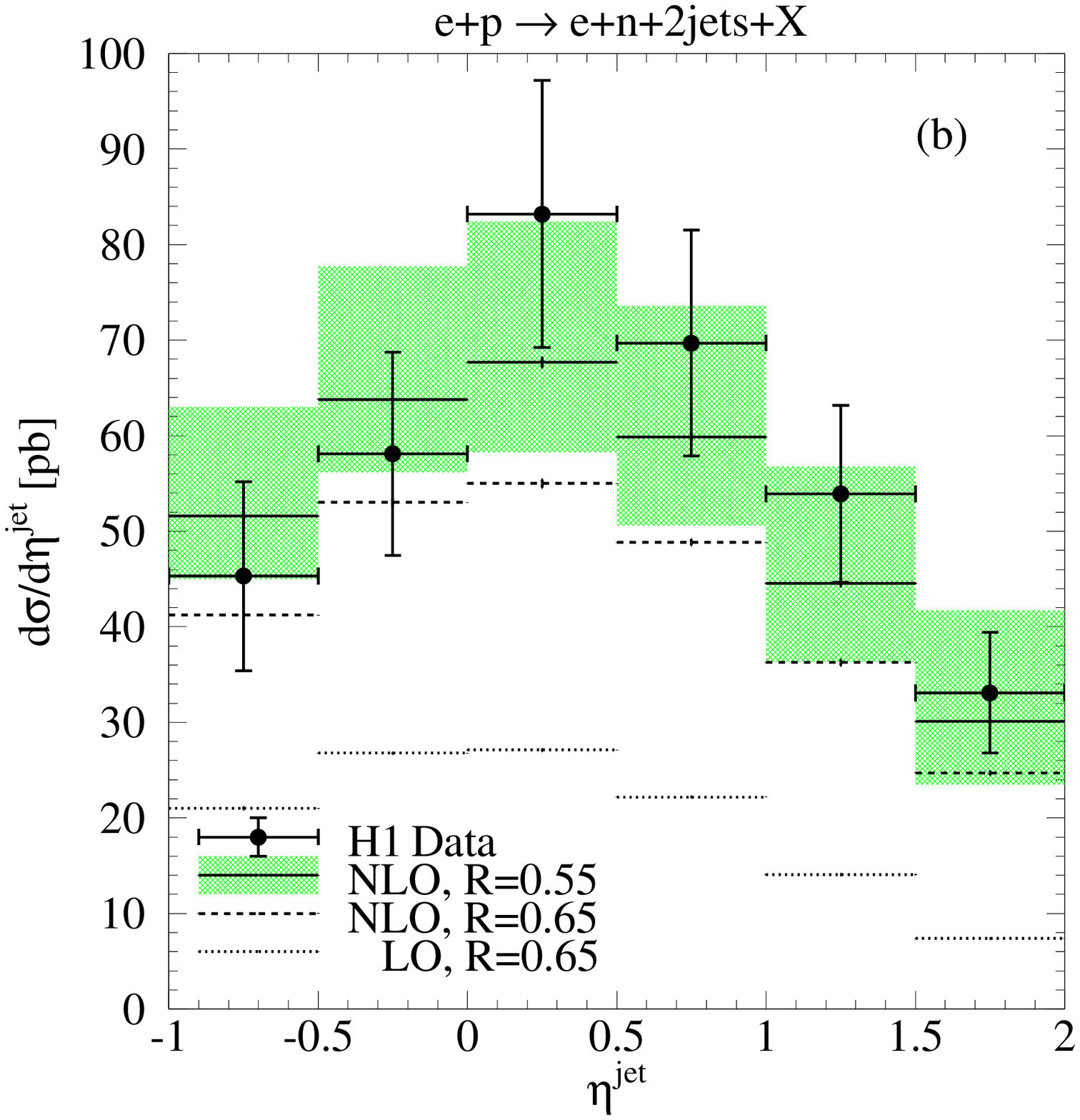}
 \includegraphics[width=0.49\columnwidth]{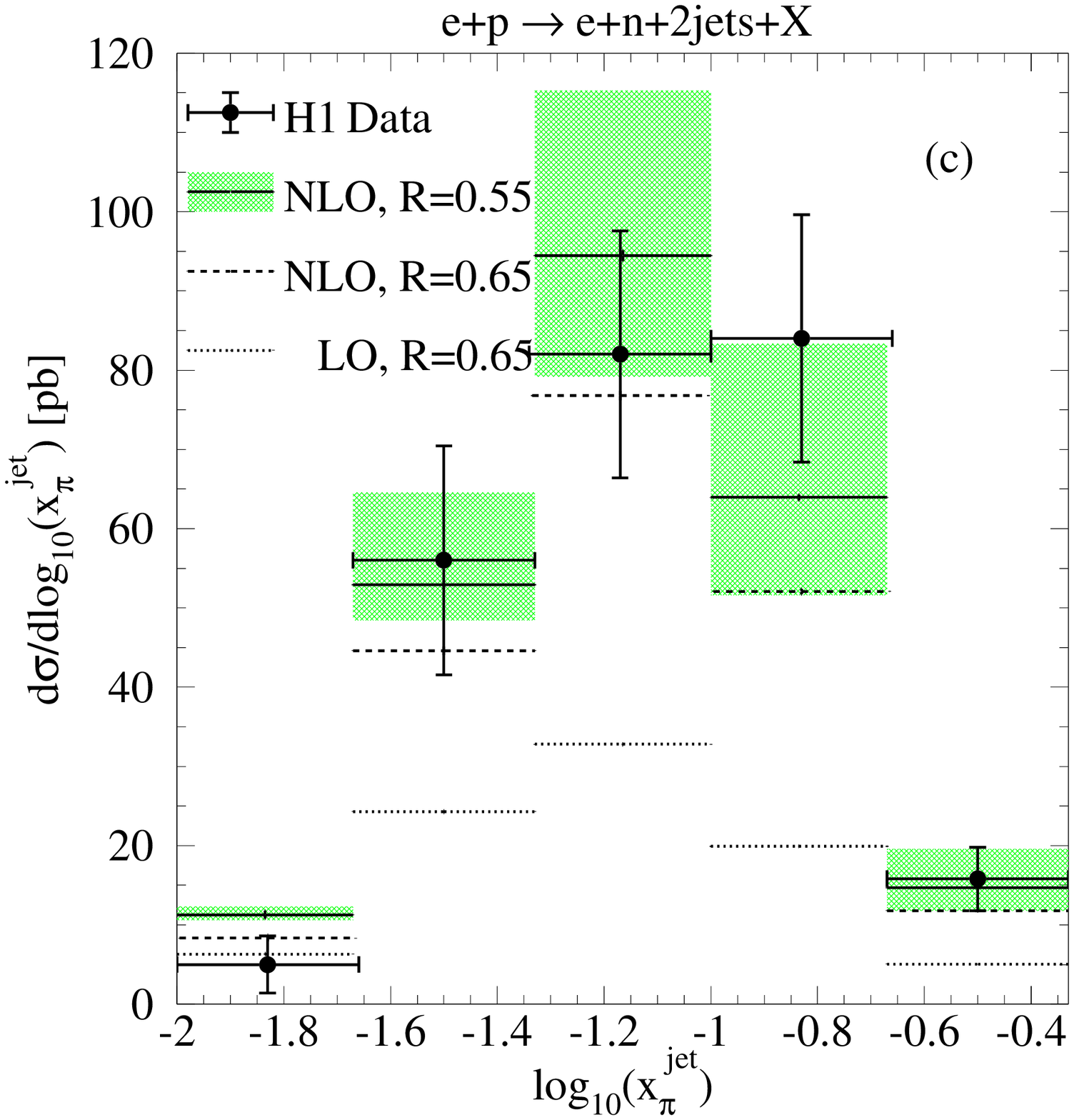}
 \includegraphics[width=0.49\columnwidth]{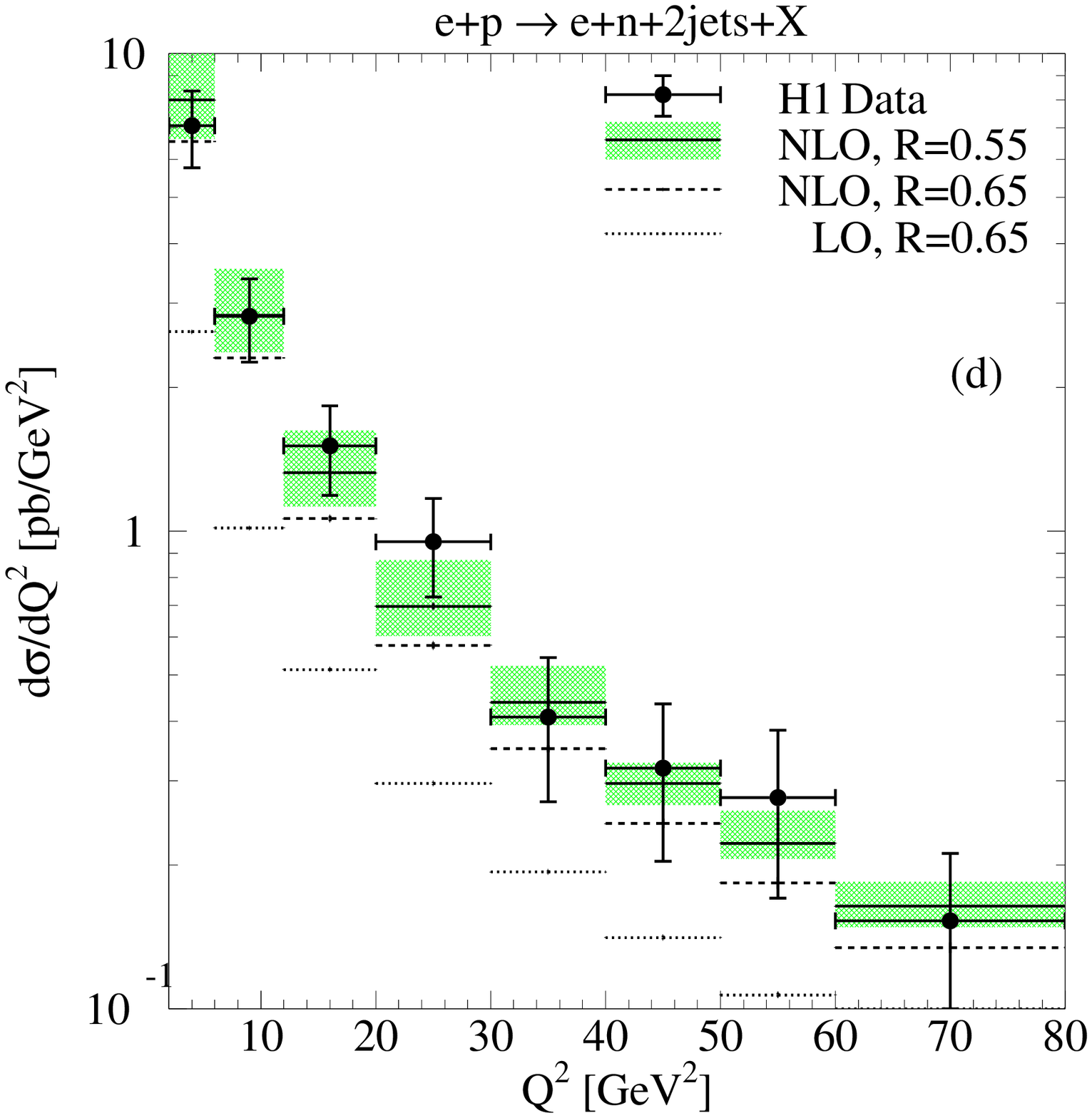}
 \caption{\label{fig:2}Differential cross sections for deep-inelastic dijet
 production with a leading neutron. The H1 data are compared to perturbative
 QCD predictions in LO and NLO and with two different values for the
 pion-neutron radius $R$.}
\end{figure*}
%
In these figures we have plotted the experimental data from \cite{10} and 
three theoretical predictions, in LO and NLO with $R = 0.65$
GeV$^{-1}$ and one in NLO 
with $R = 0.55$ GeV$^{-1}$. For $R =0.55$ GeV$^{-1}$ we show also the scale variation in NLO by 
varying the scale in the interval $\mu/2$ to $2\mu$. It is seen quite clearly 
that the NLO prediction with $R = 0.65$ GeV$^{-1}$ does not describe the data, whereas 
the prediction with $R = 0.55$ GeV$^{-1}$ is in reasonable agreement with the measured 
cross sections. We can quantify these statements by calculating the 
$\chi^2_{DF}$ for the two cases. For $R = 0.65~(0.55)$ GeV$^{-1}$ we get 
$\chi^2_{DF}=1.78~(0.78)$ for 24 degrees of freedom. We observe that the 
NLO corrections are very important in all four distributions, as one can see 
by comparing the predictions for $R = 0.65$ GeV$^{-1}$ in LO and NLO. Our theoretical 
predictions will be modified by hadronic corrections, which stand for the 
difference between jets from hadrons, on which the measured cross sections 
rely, and jets built out of partons which are the subject of the theoretical
predictions. Unfortunately these hadronic corrections, which are usually
obtained from Monte Carlo programs, which simulate LO cross sections with
parton showering and parton-to-hadron transitions, are not available for the
DIS dijet cross sections with tagged neutrons. From experience with such
cross sections with tagged protons, we know that these corrections are of
the order of $10-20\%$ except for the $x_\gamma^{\rm jet}$-distribution,
which is very much changed by these corrections. Actually, $\d\sigma/
\d x_{\gamma}$ has also been measured in \cite{10} for DIS dijets. Due
to the sensitivity of this distribution to hadronic corrections, we
did not consider this cross section suitable to give us the a
trustworthy $R$-value.

\section{Photoproduction Cross Sections}
\label{sec:3}

As in \cite{8}, the calculation of the photoproduction cross sections
is based on the formalism fully described in our previous work \cite{23}.
The cross sections which we shall evaluate are the same as in the DIS
case described above. However, we now include also $\d\sigma/\d
x_{\gamma}^{\rm jet}$ in order to see the contribution of the 
resolved part in the region $x_{\gamma}^{\rm jet} < 1$ more clearly. In the
following we show the results for $\d\sigma/\d E_T^{\rm jet}$,
$\d\sigma/\d\eta^{\rm jet}$, $\d\sigma/\d x_{\gamma}^{\rm jet}$
and $\d\sigma/\d\log_{10} (x_{\pi}^{\rm jet})$ in Figs.\ 3a-d. In
these figures, four different predictions are plotted and compared
%
\begin{figure*}
 \centering
 \includegraphics[width=0.49\textwidth]{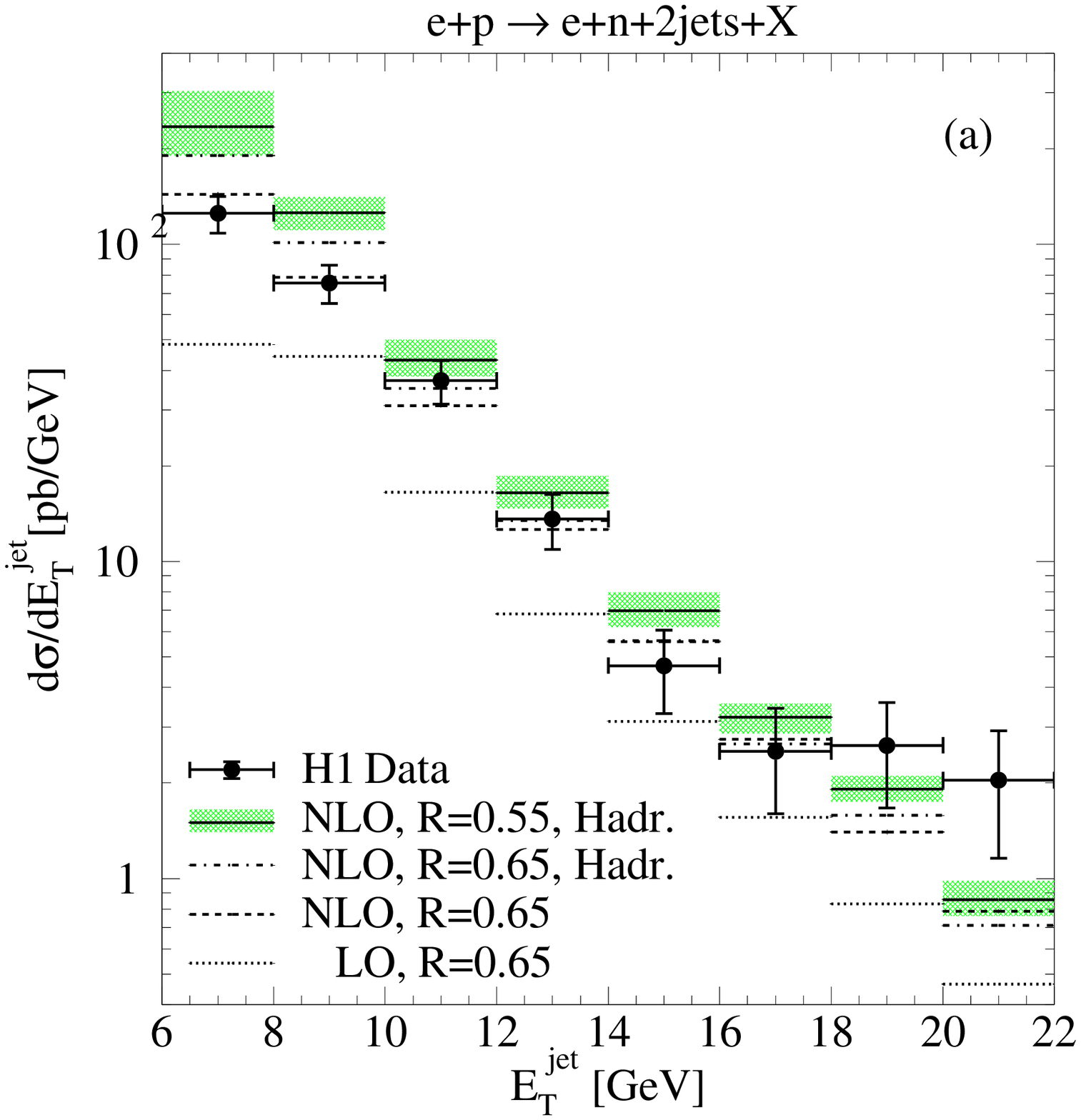}
 \includegraphics[width=0.49\textwidth]{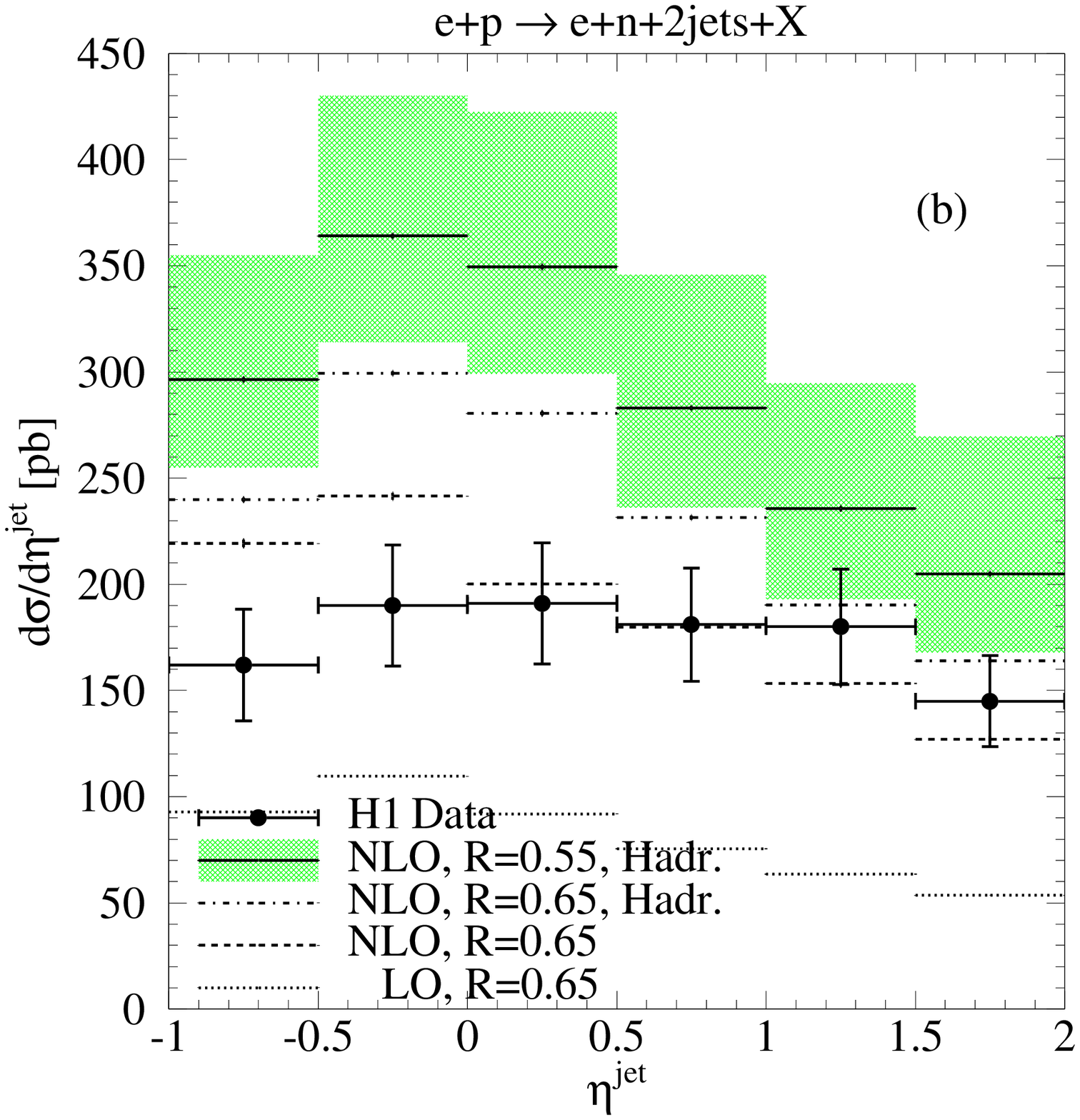}
 \includegraphics[width=0.49\textwidth]{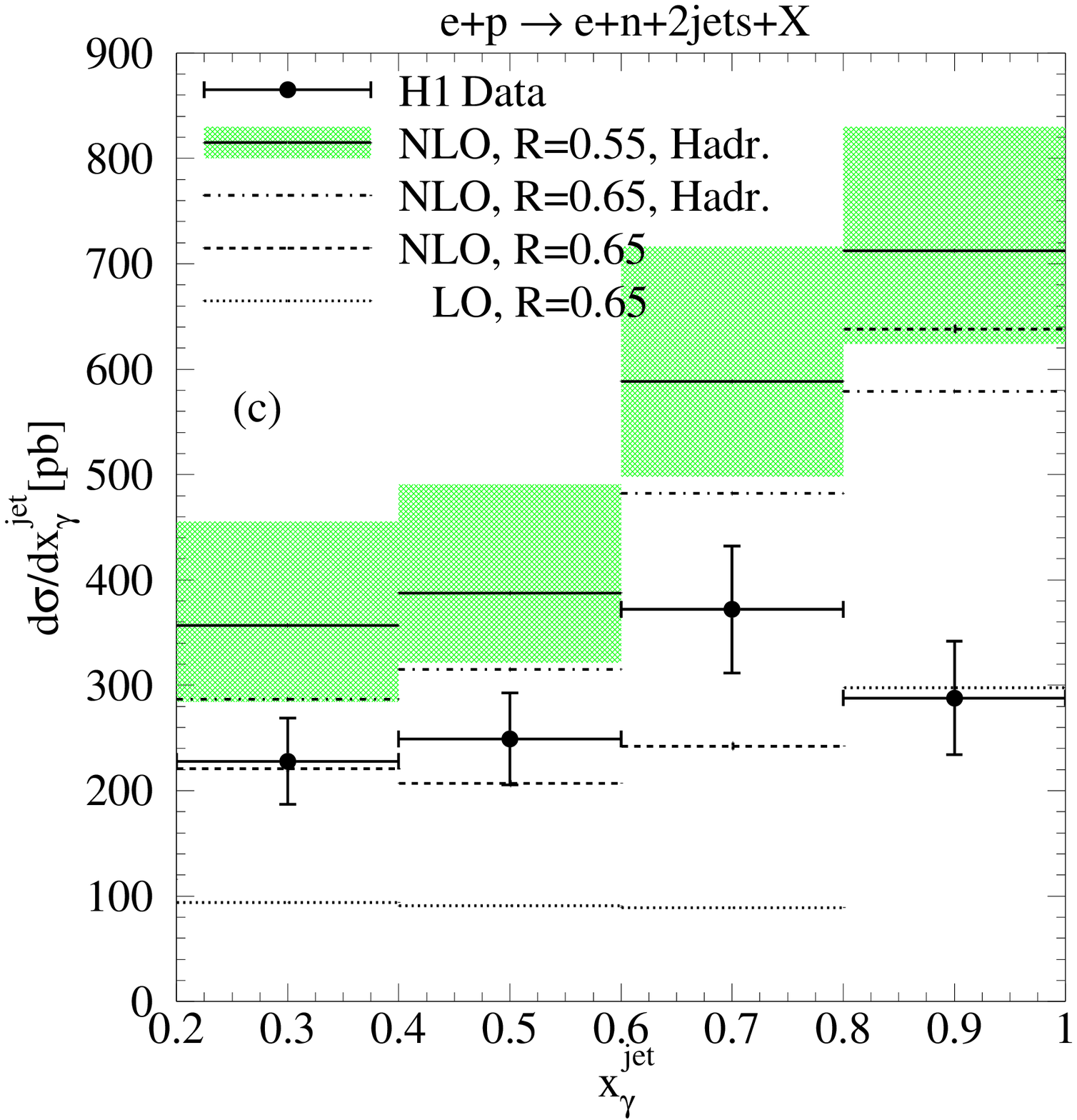}
 \includegraphics[width=0.49\textwidth]{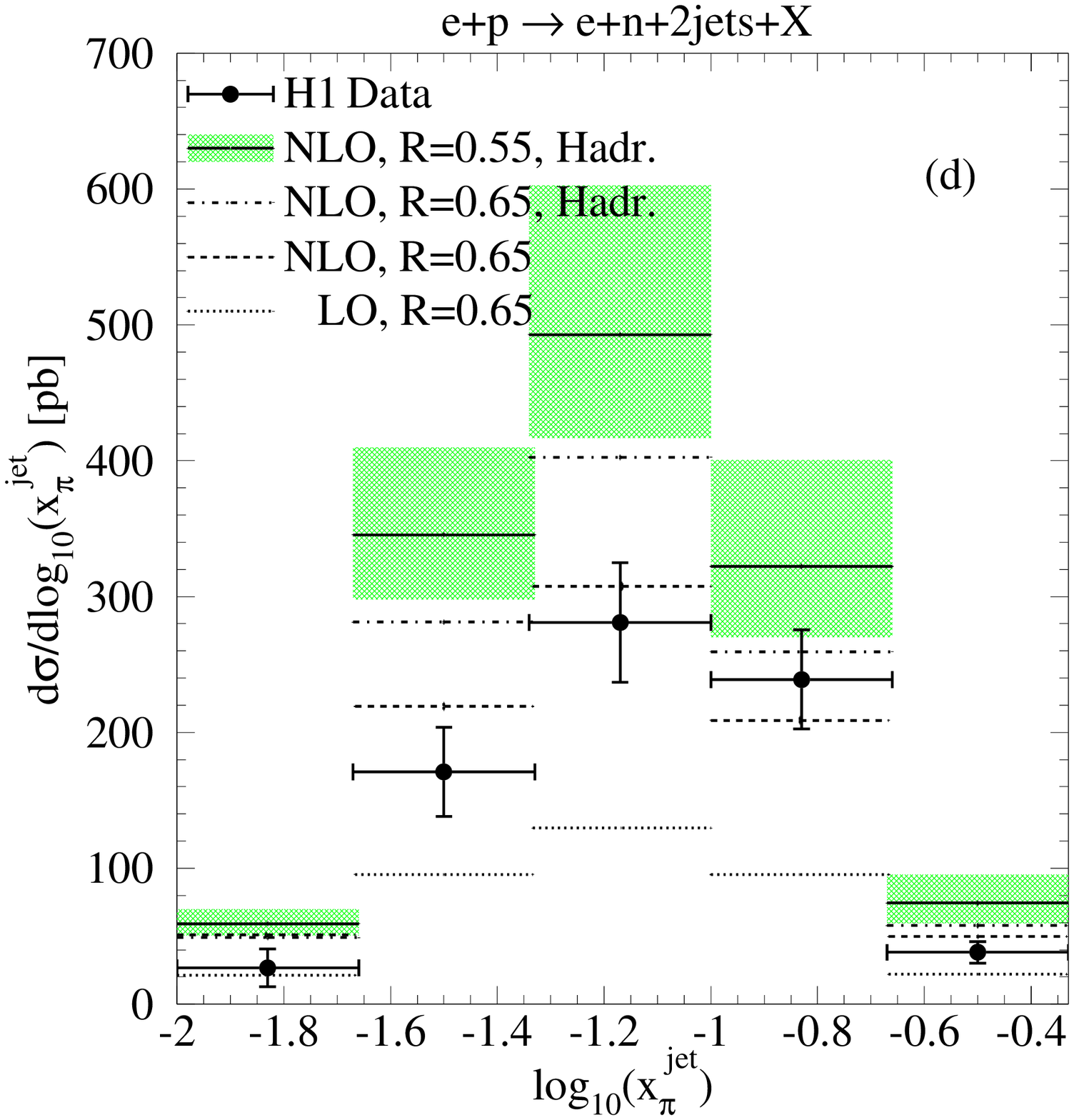}
 \caption{\label{fig:3}Differential cross sections for dijet photoproduction
 with a leading neutron. The H1 data are compared to perturbative
 QCD predictions in LO and NLO and with two different values for the
 pion-neutron radius $R$.}
\end{figure*}
%
with the data from \cite{10}. Three predictions are in NLO, one with the radius
$R=0.55$ GeV$^{-1}$, where hadronic corrections \cite{10,24} are included, and two with
$R=0.65$ GeV$^{-1}$ with and without hadronic corrections, and one is in LO with $R=0.65$ GeV$^{-1}$
without hadronic corrections. The LO prediction is far off the experimental
data showing that only NLO predictions are relevant and that the
$K$-factor ($K=$ NLO/LO) is large. For the $E_T^{\rm jet}$-distribution
in Fig. 3a,
the prediction with $R=0.55$ GeV$^{-1}$, where the shaded band gives the scale-dependence
of the cross sections, lies always above the data points except for the two
largest $E_T$-bins. The prediction with $R=0.65$ GeV$^{-1}$ agrees much better with the 
data in agreement with the results of \cite{10}. In \cite{10} the cross 
sections are somewhat smaller, since there $\alpha_s$ is smaller due
to the choice of
a smaller $\Lambda^{(4)}_{\overline{MS}}$ value ($0.200$ GeV). From this
figure, it is clear already that the radius $R=0.55$ GeV$^{-1}$ gives a bad description
of the H1 data and a reasonable agreement would be possible only for 
$R \geq 0.65$ GeV$^{-1}$. This is seen even more clearly in Figs. 3b,c,d, where the
$\eta^{\rm jet}$, $x_{\gamma}^{\rm jet}$ and $\log_{10}(x_{\pi}^{\rm jet})$
distributions are compared to the measured cross sections in \cite{10}. The
predictions lie above the data points, even if the theoretical error
estimated by the scale dependence is taken into account. To reproduce the data 
in Figs.\ 3a-d a radius larger than $R=0.65$ GeV$^{-1}$ would be needed. However,
the radius $R=0.55$ GeV$^{-1}$
was fixed by the DIS dijet cross sections. Our results in Figs. 3a-d, when
compared to the photoproduction data of the H1 collaboration, demonstrate 
that they can not be explained with the same pion flux plus pion PDFs. i.e.\
with $R=0.55$ GeV$^{-1}$. This shows us that factorization breaking is present in
photoproduction of dijets with a leading neutron
with a breaking factor of $S\simeq 0.6$. This
breaking factor changes somewhat from cross section to cross section and as a
function of the kinematic variables.

The details are shown in Figs.\ 4a-d,
%
\begin{figure*}
 \centering
 \includegraphics[width=0.49\textwidth]{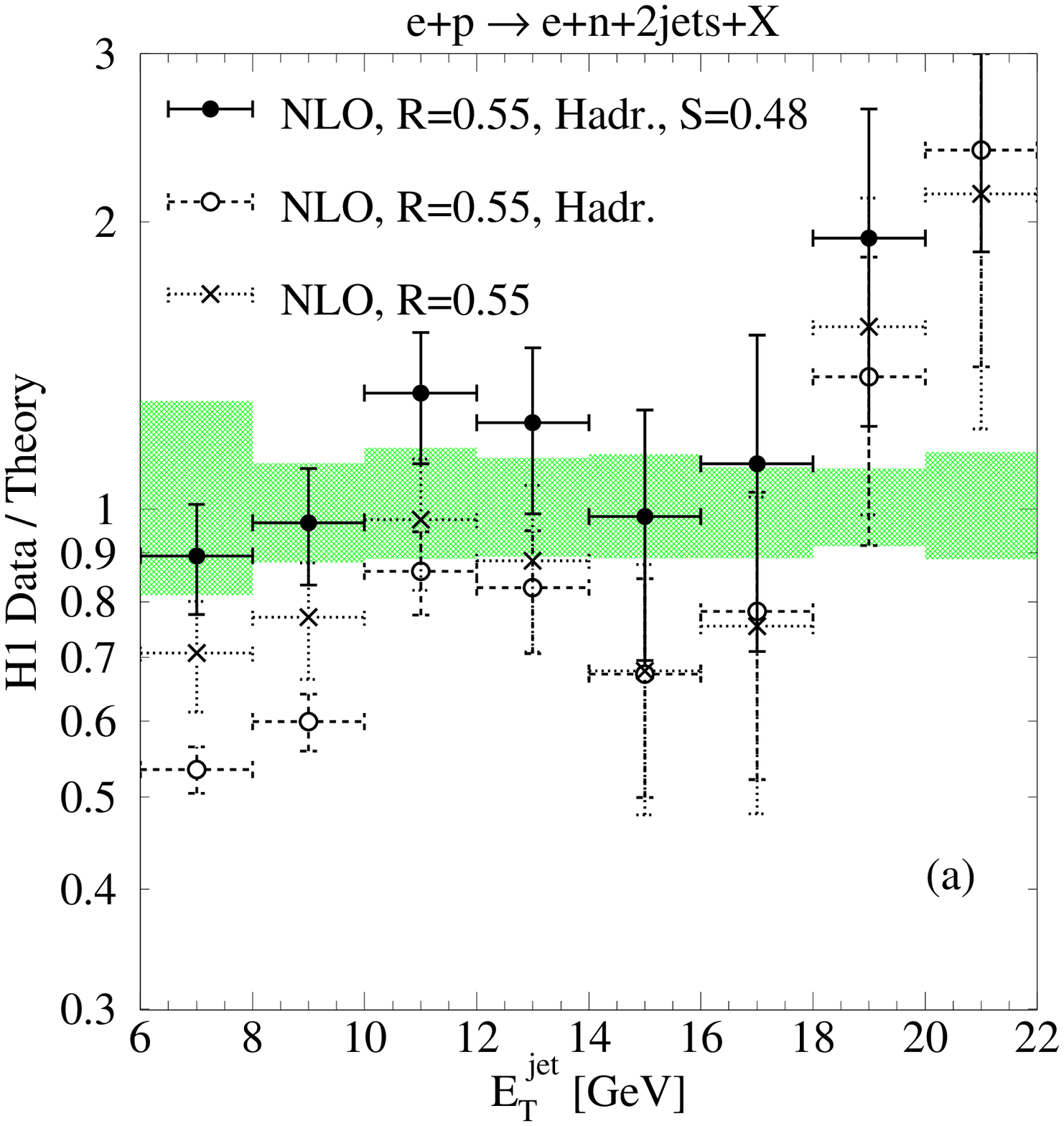}
 \includegraphics[width=0.49\textwidth]{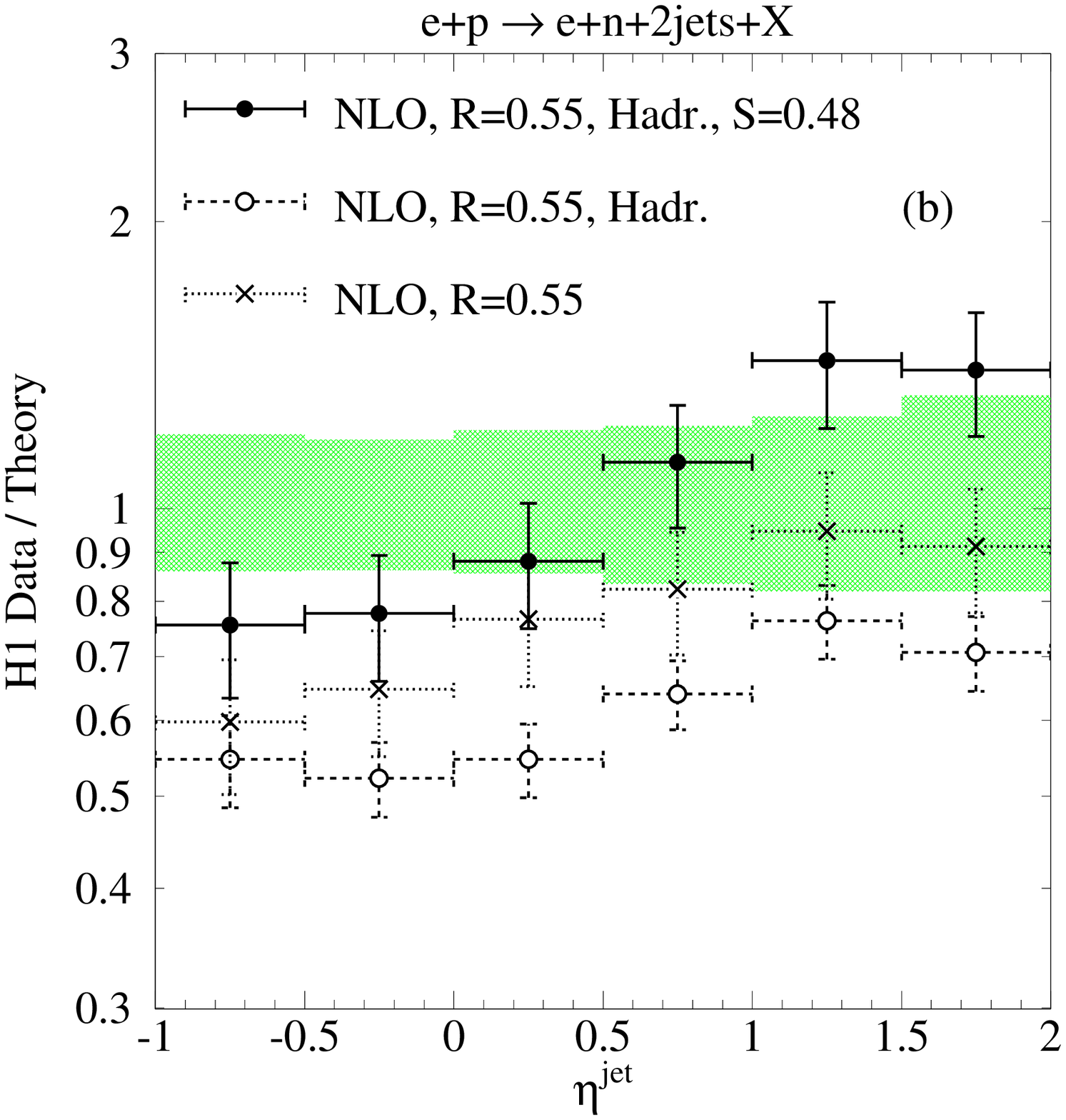}
 \includegraphics[width=0.49\textwidth]{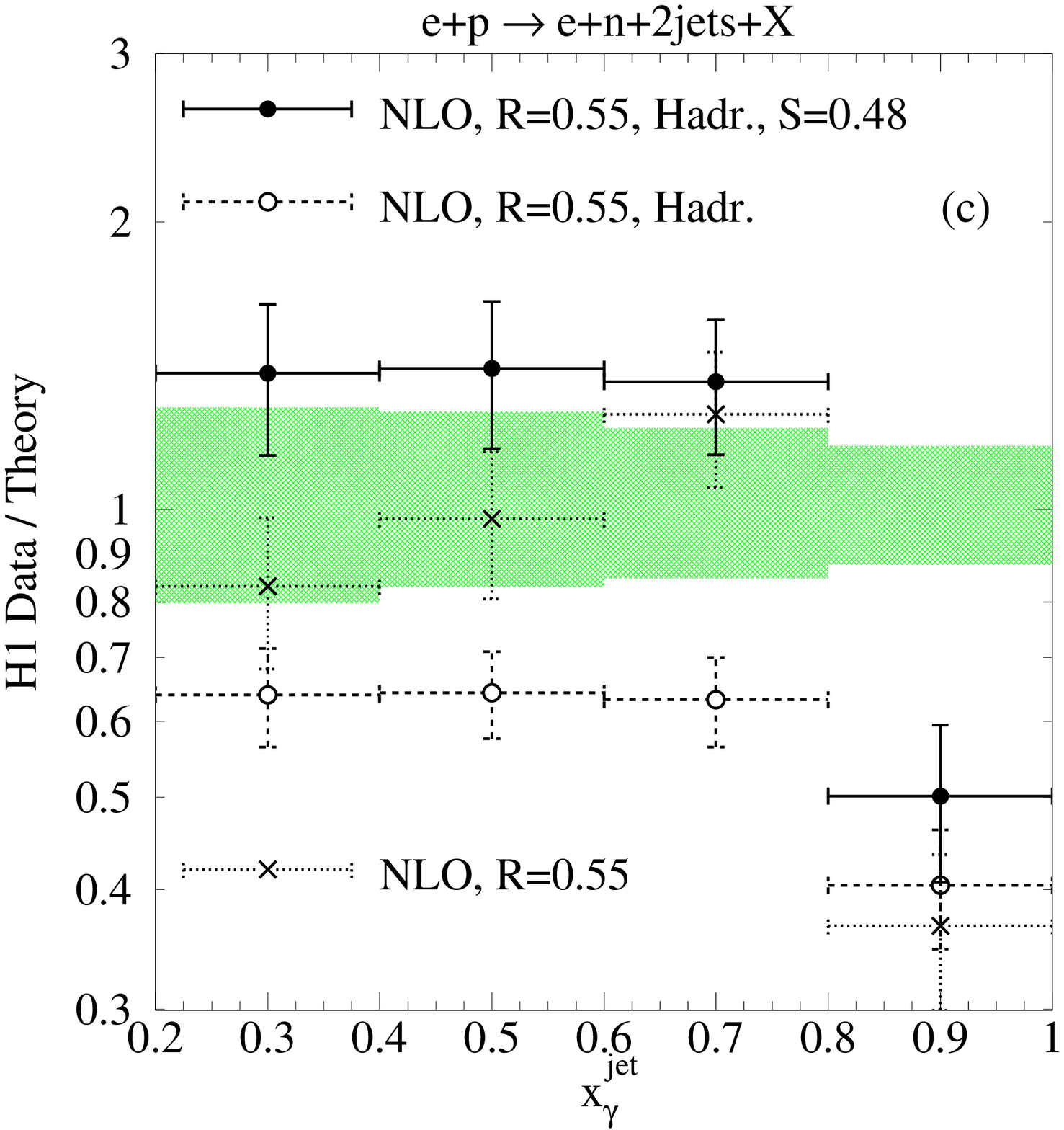}
 \includegraphics[width=0.49\textwidth]{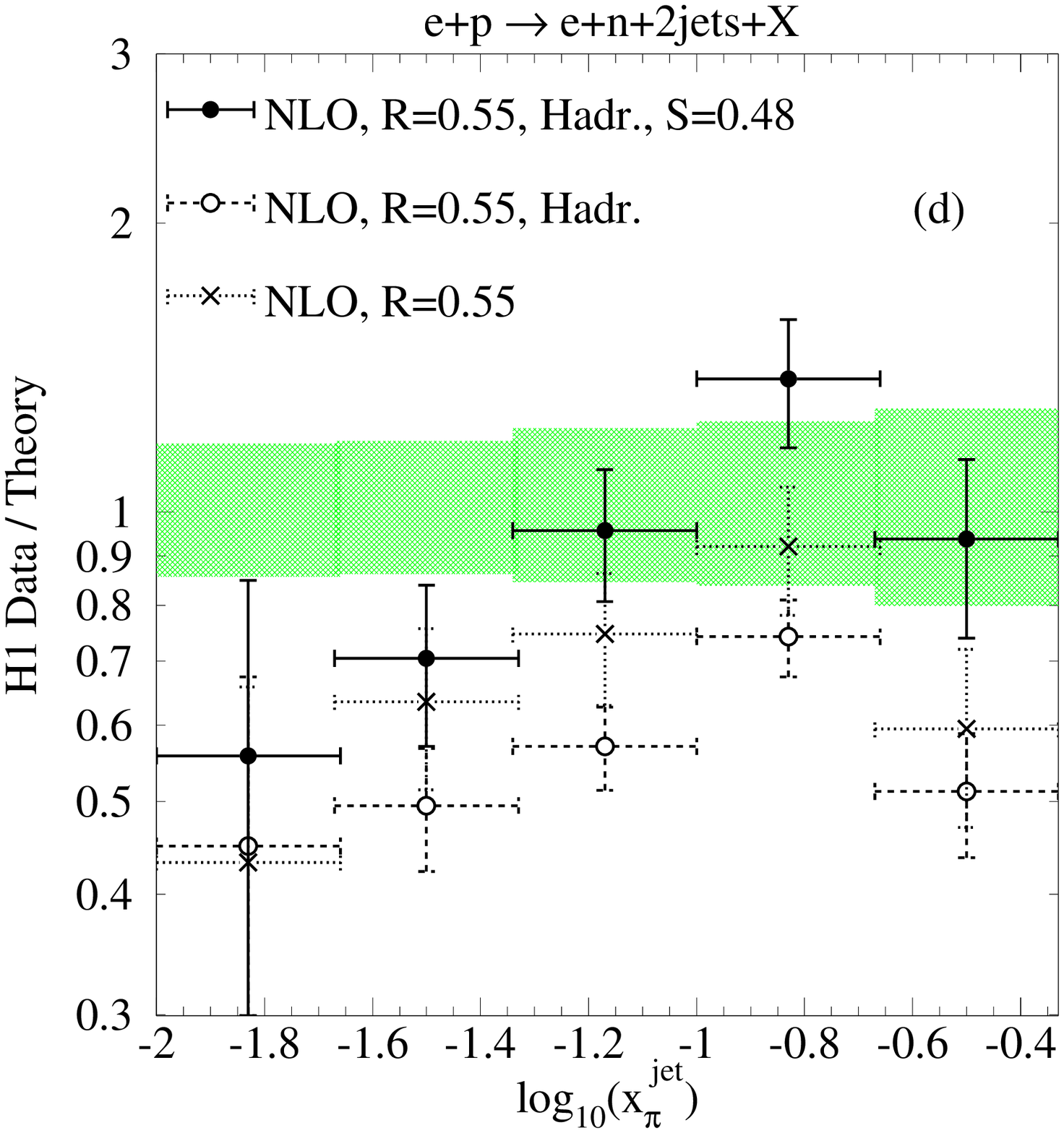}
 \caption{\label{fig:4}Ratios of H1 Data over NLO QCD for dijet photoproduction
 with a leading neutron, with and without hadronization corrections and
 including (full points/lines) a suppression factor $S$ for the resolved
 component and its scale-dependent direct NLO counterpart.}
\end{figure*}
%
where we show the ratio H1 Data/Theory for the four cross sections with
$R=0.55$ GeV$^{-1}$, with and without hadronic corrections. In addition to the ratios 
following from the results in Figs.\ 3a-d, we show in Figs.\ 4a-d also another
ratio, where the theoretical prediction is again obtained with $R=0.55$ GeV$^{-1}$
including hadronic corrections, but now with a suppression factor $S=0.48$
applied to the resolved cross section together with the $\mu_{\gamma}$
scale-dependent part
of the NLO corrections to the direct cross section. Here, $\mu_{\gamma}$ is the
factorization scale of the photon vertex. This additional suppression in the
direct cross section introduced in \cite{25} serves the purpose to eliminate 
the $\mu_{\gamma}$-dependence of the sum of the direct and resolved cross 
section. We see from Figs.\ 4a-d that this ratio lies near to one if we 
take into account the scale variation of the theoretical prediction. The
$x_{\gamma}^{\rm jet}$-distribution yields values for this ratio larger than one (except
for the last bin), which overlap, however, with the scale variation band. It
is known that the $x_{\gamma}^{\rm jet}$-distribution suffers from rather large hadronic
corrections as seen for example Fig.\ 3c, where the correction in the bin
$x_{\gamma}^{\rm jet} \in [0.6;0.8]$ is as large as a factor of two.

The two suppression factors, $S\simeq 0.6$ for the full direct and
resolved contributions or $S=0.48$ for the resolved and initial-state
singular part of the direct contribution, can be compared with the
suppression factor obtained in \cite{28}. In this work, the spectra of
leading neutrons, both in photoproduction and in DIS, were studied and
compared to recent ZEUS experimental data \cite{14}. It was found that
the photoproduction cross section on the basis of the pion-exchange
model agreed with the data if it was reduced by a factor of about
0.4. This factor could be quite well explained by absorptive
corrections to pion exchange and can be compared with the suppression
factor of 0.48 for the resolved/direct-IS components. Furthermore, it
must be emphasized that the absorptive corrections for the total
photoproduction cross section with a tagged neutron may differ from
the absorptive corrections for the production of a pair of high-$E_T$
jets, even in the resolved case.

In Fig.\ 5, we show separately the $E_T^{\rm jet}$-distribution
%
\begin{figure}
 \centering
 \includegraphics[width=\columnwidth]{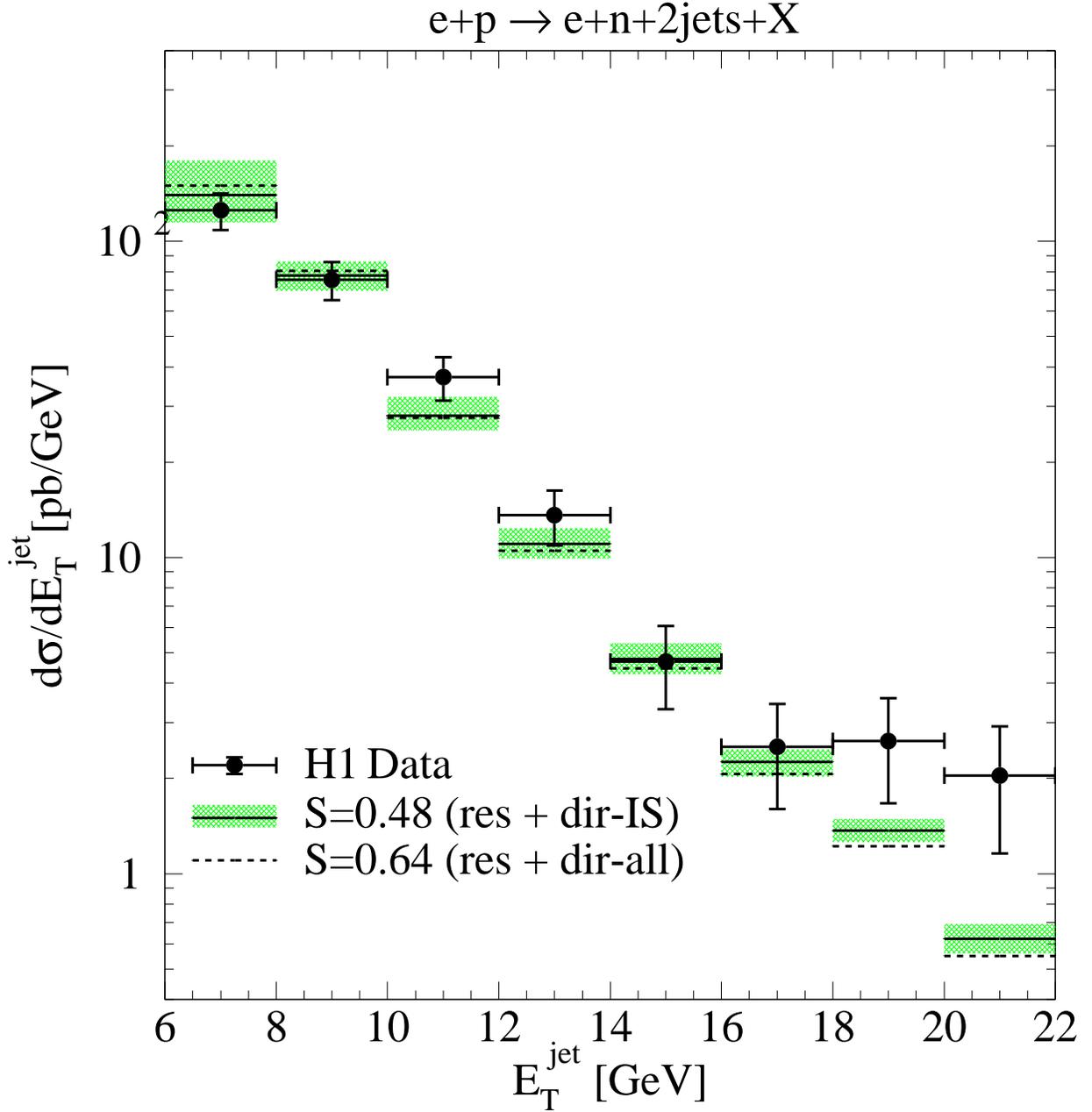}
 \caption{\label{fig:5}$E_T^{\rm jet}$-distribution of photoproduced dijets with a
 leading neutron. The H1 data are compared to NLO QCD calculations corrected
 for hadronization effects in the pion-exchange model with pion-neutron radius
 $R=0.55$ GeV$^{-1}$ and suppression of the resolved/direct initial-state contributions
 (full) and of all direct and resolved photon contributions (dashed).}
\end{figure}
%
as measured by the H1 collaboration and compare it to theoretical NLO QCD
predictions including hadronization corrections and using the radius $R=0.55$ GeV$^{-1}$.
In the full histogram, we test the hypothesis of suppressing only the resolved
(and direct initial-state) contributions with a factor $S=0.48$
and obtain quite a reasonable value of $\chi^2_{DF}=1.2$. The
shaded band indicates again the theoretical uncertainty coming from
simultaneous variation of the renormalization and factorization scales by
a factor of two around the central scale. It overlaps with the experimental
error bars in all but the highest two bins, whose precision is obviously
limited by statistics. In contrast, the dashed histogram tests the hypothesis
of suppressing all direct and resolved contributions with a fitted factor $S=0.64$,
as proposed in the H1 analysis of diffractive dijet photoproduction, leading
to a higher value of $\chi^2_{DF}=1.7$. We take this as an indication
that the first hypothesis describes this most robust (and exponentially
falling) distribution better.

Finally, we wish to comment on our earlier analysis \cite{8} of the
photoproduction dijet cross sections as measured by the ZEUS
collaboration several years ago \cite{9}. At this time, data on dijets
with tagged neutrons in DIS were not available. Therefore, we fitted
the pion-neutron radius $R$ of the light-cone form factor to the ZEUS
data with the result $R=0.5$ GeV$^{-1}$. As shown in this work, the H1
photoproduction data \cite{10} are badly described with this
radius. The reason for this mismatch is the fact that in \cite{8} we
have chosen different parameterizations of the pion PDFs, i.e.\ SMRS
versus GRV \cite{11}, and of the photon PDFs, i.e.\ GS96 \cite{29}
versus GRV \cite{19}. In particular, the choice of the pion PDFs has a
strong influence on the absolute value of the cross section (see for
example Fig.\ 3 in \cite{8}). The ZEUS collaboration demonstrated in
2002 \cite{14} that the shape of their measured $F_2^\pi$ distribution
agrees quite well with the GRV parameterization and much less with
SMRS. Furthermore, the normalization of the theoretical dijet cross section
was influenced by the choice of the difference in the $E_T$-cuts for
the two jets, which was set to zero in the analysis of the
experimental data.

\section{Conclusion}
\label{sec:4}

In Summary, we have performed a comprehensive NLO QCD analyis of dijet
production with a leading neutron in both deep-inelastic scattering
and photoproduction. We emphasized the question whether factorization
breaking occurs not only in diffractive photoproduction, but also in
photoproduction with $p\to n$ transitions. Assuming that the latter
are dominated by one-pion exchange and well described by GRV pion
PDFs and a light-cone flux factor, we were able to fit its free parameter,
the pion-neutron radius $R$, to recent H1 DIS data with the result
$R=0.55$ GeV$^{-1}$ for $\chi^2_{DF}=0.78$ and 24 degrees of freedom.
In contrast, a radius of $R=0.65$ GeV$^{-1}$ as used by the H1
collaboration leads to a considerably larger $\chi^2_{DF}$ of 1.78.

When applying the fitted radius of $R=0.55$ GeV$^{-1}$ to our NLO QCD
predictions for dijet photoproduction with a leading neutron, we
seriously overestimate the corresponding H1 data. We therefore
conclude that factorization
breaking occurs not only in diffractive $ep\to e'pX$, but also in
$ep\to e'nX$ scattering. Only after including absorptive
corrections, a good phenomenological description of the H1 data
can be obtained. A suppression of resolved photoproduction and its
factorization-scale dependent direct NLO counterpart only with a
suppression factor of $S=0.48$ seems to be favored over a global
suppression of all direct and resolved contributions with a
suppression factor of $S=0.64$, at least in the
steeply falling $E_T^{\rm jet}$-distribution, where we obtain values
of $\chi^2_{DF}=1.2$ versus 1.7.

\begin{acknowledgement}
M.K.\ gratefully acknowledges the hospitality of the Institute of
Theoretical Physics at the University of G\"ottingen, where part of
this work has been completed.
\end{acknowledgement}



\end{document}